\documentstyle[12pt]{article}
\newcommand{\ret}{\nonumber \\}
\newcommand{\Section}[1]%
{\section{#1}\setcounter{equation}{0}%
\setcounter{theorem}{0}}
\newtheorem{theorem}{Theorem}
\newtheorem{lemma}[theorem]{Lemma}

\newtheorem{definition}[theorem]{Definition}

\newenvironment{proof}[1]%
{\par\noindent{\em #1:\ }}%
{~\rule{2mm}{2mm}\par\bigskip}
\begin{document}
\newpage\thispagestyle{empty}
{\topskip 2cm
\begin{center}
{\large\bf Spectral Gap and Decay of Correlations\\} 
\bigskip
{\large\bf in U(1)-Symmetric Lattice Systems in Dimensions $D<2$\\} 
\bigskip\bigskip\bigskip\bigskip
{\large Tohru Koma}\\
\bigskip
{\small \it Department of Physics, Gakushuin University, 
Mejiro, Toshima-ku, Tokyo 171-8588, JAPAN}\\
\smallskip
{\small\tt e-mail: tohru.koma@gakushuin.ac.jp}
\end{center}
\vfil
\noindent
We consider many-body systems with a global U(1) symmetry 
on a class of lattices with the (fractal) dimensions $D<2$ 
and their zero temperature correlations whose observables behave 
as a vector under the U(1) rotation. For a wide class of the models, 
we prove that if there exists a spectral gap above the ground state, 
then the correlation functions have a stretched exponentially decaying 
upper bound. This is an extension of the McBryan-Spencer method 
at finite temperatures to zero temperature. The class includes quantum spin 
and electron models on the lattices, and our method also allows finite 
or infinite (quasi)degeneracy of the ground state. The resulting bounds 
rule out the possibility of the corresponding magnetic 
and electric long-range order.  
\par\noindent
\bigskip
\hfill
\vfil}\newpage
\Section{Introduction}

As is well known, low dimensional systems show large fluctuations 
for continuous symmetry. 
The most famous result is the Hohenberg-Mermin-Wagner theorem~\cite{HMW}
which states that the corresponding spontaneous magnetizations are 
vanishing at finite temperatures in one and two dimensions. 
Since their articles appeared, their method have been applied to 
various systems~\cite{HMW2} including classical and quantum magnets, 
interacting electrons in a metal and Bose gas.\footnote{For a mathematically 
rigorous treatment for the unbounded operators, 
see ref.~\cite{BM}.} The theorem was extended to 
the models on a class of generic lattices with the fractal dimensions 
$D\le 2$ by Cassi \cite{Cassi}.
In a stronger sense, it was also proved for a class of low-dimensional 
systems that the equilibrium states are invariant under the action of 
the continuous symmetry group \cite{FP,BFK}.
Even at zero temperature, the same is true~\cite{PS,Shastry,NS} 
if the corresponding one- or two-dimensional system 
satisfies conditions~\cite{Momoi} such as boundedness of  
susceptibilities.\footnote{Since a single spin shows the spontaneous 
magnetization at zero temperature, the absence of the spontaneous 
symmetry breaking implies that the strong fluctuations due to 
the interaction destroy the ordering and lead to the finite susceptibilities. 
In other words, one cannot expect the absence of spontaneous symmetry breaking 
at zero temperature in a generic situation.} 

About the corresponding long range correlations, 
Fisher and Jasnow \cite{JF} proved clustering properties 
of two point functions by using the Bogoliubov 
inequality. See also ref.~\cite{BFK}.  
McBryan and Spencer \cite{MS} obtained a better decay for two point 
correlations of classical spin systems. Their method has been applied 
various classical and quantum systems. 
The resulting upper bounds for the correlations decay by power, 
exponential or stretched exponential laws \cite{Ito,Picco,vanEnter,KT1,MR,KT2}
and rule out the ordering at finite temperatures in (fractal) dimensions $D\le 2$. 
But a zero temperature analogue of the McBryan-Spencer bound has 
not yet been obtained. 

On the other hand, since Haldane~\cite{Haldane} predicted a ``massive" phase in 
low dimensional, isotropic quantum systems, 
many examples have been found to have 
a spectral gap above the ground state and exponentially decaying correlations 
in the ground state as initiated by \cite{AKLT}. 
See also related articles \cite{MG,Kennedy,FNW}. 
These examples raise the following question: 
Consider a generic low dimensional system with a continuous symmetry 
and its zero temperature correlations whose observables show a non-trivial  
representation under the action of the continuous symmetry group. Then a spectral gap 
above the ground state implies (stretched) exponential decay of the correlations? 
And if so, the corresponding upper bound can be obtained 
by the McBryan-Spencer method?  
   
In this paper, we address this problem. We consider the correlation functions 
whose observables behave as a vector under a U(1) rotation. 
In order to estimate the decay of the correlation functions,  
we extend the McBryan-Spencer method to zero temperature 
under the assumption that there exists a spectral gap above the ground state.  
As a result, we prove that the correlation functions have 
a stretched exponentially decaying upper bound. 
This method covers a wide class of many-body systems with a global U(1) symmetry 
on a class of lattices with the (fractal) dimensions $D<2$. 
We stress that this method is also an extension of the Combes-Thomas
method~\cite{CT} for Schr\"odinger operators to many-body systems in statistical 
mechanics. In the next section, we will give the precise definition of 
the class of the fractal lattices and describe the results 
for two typical examples, the Heisenberg and the Hubbard models. 
The proof of the main results will be given in Section~\ref{proof}.

\Section{Models and results}
\label{model}

We begin with defining the class of the (fractal) lattices which we consider 
in this paper. The class of the lattices is the same as in~\cite{KT2}. 
See also \cite{Fractal} for fractal lattices and models on the lattices.  

Consider first a connected lattice $\Lambda=(\Lambda_s,\Lambda_b)$, 
where $\Lambda_s$ is a set of sites, $i,j,k,\ell,\ldots$, and 
$\Lambda_b$ is a set of bonds, i.e., pairs of sites, 
$\{i,j\},\{k,\ell\},\ldots$. If a sequence of sites, $i_0,i_1,i_2,\ldots,i_r$, 
satisfies $\{i_{n-1},i_n\}\in\Lambda_b$ for $n=1,2,\ldots,r$, then 
we say that the path, $\{i_0,i_1,i_2,\ldots,i_r\}$, has length $r$ and 
connects $i_0$ to $i_r$.  
We define the ``sphere", $S_r(m)$, centered at $m\in \Lambda_s$ 
with the radius $r$ as  
\begin{equation}
S_r(m):=\{\ell\in\Lambda_s|{\rm dist}(\ell,m)=r\},
\end{equation}
where ${\rm dist}(\ell,m)$ is the graph-theoretic distance 
which is defined to be the shortest path length that one needs to connect $\ell$ to $m$. 
Let $|A|$ denote the number of the elements in the set $A$.
We assume that there exists 
a ``(fractal) dimension" $D\ge 1$ of the lattice $\Lambda$ such that 
the number $|S_r(m)|$ of the sites in the sphere satisfies
\begin{equation}
\sup_{m\in\Lambda_s}|S_r(m)|\le C_0r^{D-1}
\label{dimensionD}
\end{equation}
with some positive constant $C_0$. 

We consider spin or fermion systems with a global U(1) symmetry 
on the lattice $\Lambda$ with the (fractal) dimensions $1\le D<2$.
We require the existence of a ``uniform gap" above the sector of the 
ground state of the Hamiltonian $H_\Lambda$. 
The precise definition of the ``uniform gap" is:  

\begin{definition}
\label{definition}
We say that there is a uniform gap above the sector of the ground state
if the spectrum $\sigma(H_\Lambda)$ of the Hamiltonian $H_\Lambda$ 
satisfies the following conditions: 
The ground state of the Hamiltonian $H_\Lambda$ is 
$q$-fold (quasi)degenerate in the sense that there are $q$ eigenvalues, 
$E_{0,1},\ldots, E_{0,q}$, in the sector of the ground state 
at the bottom of the spectrum of $H_\Lambda$ such that 
\begin{equation}
\Delta{\cal E}:=\max_{\mu,\mu'}\{|E_{0,\mu}-E_{0,\mu'}|\}\rightarrow 0\quad\mbox{as }\
|\Lambda_s|\rightarrow\infty.
\label{defDeltacalE} 
\end{equation}
Further the distance between the spectrum, $\{E_{0,1},\ldots, E_{0,q}\}$,   
of the ground state and the rest of the spectrum is larger than 
a positive constant $\Delta E$ which is independent of the volume $|\Lambda_s|$. 
Namely there is a spectral gap $\Delta E$ above the sector of the ground state. 
\end{definition}
 
\noindent
{\bf Remark:} For the special case with $q=1$, the ground state is unique. 
As well known examples with $q\ne 1$, Majumdar-Ghosh model \cite{MG} 
shows a spectral gap above the degenerate ground state, 
and the spin-1 antiferromagnetic chain with open 
boundaries exhibits a spectral gap above 
the fourfold quasidegenerate ground state~\cite{Kennedy}.  
In the thermodynamic limit $|\Lambda_s|\rightarrow\infty$, 
we allow infinite degeneracy $q=\infty$ of the ground state as in \cite{KN}. 

Consider first a quantum spin system with a U(1) symmetry on the lattice $\Lambda$. 
As a concrete example, we consider the standard XXZ Heisenberg model 
on the lattice. The Hamiltonian $H_\Lambda$ is given by 
\begin{equation}
H_\Lambda=H_\Lambda^{XY}+V_\Lambda(\{S_i^z\})
\label{hamspin}
\end{equation}
with
\begin{equation}
H_\Lambda^{XY}=2\sum_{\{i,j\}\in\Lambda_b}J_{i,j}^{\rm XY}
(S_i^xS_j^x+S_i^yS_j^y),
\label{hamXY}
\end{equation}
where $(S_i^x,S_i^y,S_i^z)$ is the spin operator 
at the site $i\in\Lambda_s$ with the spin $S=1/2,1,3/2,\ldots$, and 
$J_{i,j}^{\rm XY}$ are real coupling constants; 
$V_\Lambda(\{S_i^z\})$ is a real function of the $z$-components, $\{S_i^z\}_{i\in\Lambda_s}$, 
of the spins.   
For simplicity, we take 
\begin{equation}
V_\Lambda(\{S_i^z\})=\sum_{\{i,j\}\in\Lambda_b}J_{i,j}^{\rm Z}S_i^zS_j^z
\end{equation}
with real coupling constants $J_{i,j}^{\rm Z}$. 
We assume that there are positive constants, $J_{\rm max}^{\rm XY}$ and 
$J_{\rm max}^{\rm Z}$, which satisfy $|J_{i,j}^{\rm XY}|\le J_{\rm max}^{\rm XY}$  
and $|J_{i,j}^{\rm Z}|\le J_{\rm max}^{\rm Z}$ for any bond $\{i,j\}\in\Lambda_b$. 
We stress that we can also treat more general interactions in the same way. 
 
Let $P_0$ be the projection onto the sector of the ground state. 
We define the ground-state expectation as 
\begin{equation}
\left\langle\cdots\right\rangle_0:=
\frac{1}{q}{\rm Tr}\ (\cdots)P_0, 
\end{equation}
where ${\rm Tr}$ stands for the trace which is over all the spin states. 
We consider the transverse spin-spin correlation, 
$\left\langle S_m^+S_n^-\right\rangle_0$, where $S_i^\pm:=S_i^x\pm i S_i^y$.

\begin{theorem}
Suppose that $1\le D<2$ and that there is a uniform gap $\Delta E$ above 
the sector of the ground state in the sense of Definition~\ref{definition}. 
Then there exists a positive constant $\gamma$ such that 
the transverse spin-spin correlation satisfies the bound,  
\begin{equation}
\left|\left\langle S_m^+S_n^-\right\rangle_0\right|\le
{\rm Const.}\exp\left[-\gamma\{{\rm dist}(m,n)\}^{1-D/2}\right], 
\end{equation}
in the thermodynamic limit $|\Lambda_s|\rightarrow\infty$. 
\end{theorem}

\noindent
{\bf Remark:} 1. It is easy to extend the result to more complicated correlations such as 
the multispin correlation 
$\left\langle S_{m_1}^+\cdots S_{m_j}^+S_{n_1}^-\cdots S_{n_j}^-\right\rangle_0$. 

\noindent
2. Recently, the exponential clustering of the correlations 
was proved for quantum many-body lattice systems by Hastings \cite{Hastings,NachSim} 
under the gap assumption. This is a non-relativistic version of 
Fredenhagen's theorem \cite{Fredenhagen} of relativistic quantum 
field theory. Combining this exponential clustering with 
the present result, a better, exponentially decaying bound 
for the ground-state correlations can be obtained \cite{HK} 
for a class of models on lattices with the (fractal) dimensions 
$D<2$ and with a certain self-similarity. The class includes 
the translationally invariant regular lattices such as ${\bf Z}$. 
\smallskip

For a finite volume $|\Lambda_s|<\infty$, the Hamiltonian $H_\Lambda$ of (\ref{hamspin}) 
commutes with $S_\Lambda^z:=\sum_{i\in\Lambda_s}S_i^z$. 
Using this symmetry, we can block diagonalize the Hamiltonian $H_\Lambda$, 
and denote by $H_{\Lambda,M}$ the restriction of $H_\Lambda$ to  
the eigenspace ${\cal H}_{\Lambda,M}$ of $S_\Lambda^z$ with the eigenvalue $M$. 
Let $P_{0,M}$ denote the projection onto the ground state of $H_{\Lambda,M}$ 
in the subspace ${\cal H}_{\Lambda,M}$. Here the ground state may be 
(quasi)degenerate in the sense of Definition~\ref{definition}. We define 
the ground-state expectation as 
\begin{equation}
\left\langle\cdots\right\rangle_{0,M}:=\frac{1}{q_M}
{\rm Tr}\ (\cdots)P_{0,M},
\end{equation}
where $q_M$ is the degeneracy of the sector of the ground state of $H_{\Lambda,M}$. 

\begin{theorem}
\label{maintheorem}
Suppose that $1\le D<2$ and that there is a uniform gap $\Delta E$ above 
the sector of the ground state of $H_{\Lambda,M}$ in the spectrum 
of $H_{\Lambda,M}$ with the eigenvalue $M$ of $S_\Lambda^z$
in the sense of Definition~\ref{definition}. 
Then there exists a positive constant $\gamma$ such that 
the transverse spin-spin correlation satisfies the bound, 
\begin{equation}
\left|\left\langle S_m^+S_n^-\right\rangle_{0,M}\right|\le
{\rm Const.}\exp\left[-\gamma\{{\rm dist}(m,n)\}^{1-D/2}\right], 
\label{SScorrbound}  
\end{equation}
in the thermodynamic limit $|\Lambda_s|\rightarrow\infty$. 
\end{theorem}

As an example of a lattice fermion system, we consider 
the following Hamiltonian \cite{KT1,MR} on the lattice $\Lambda$ 
with the (fractal) dimension $1\le D<2$: 
\begin{equation}
H_\Lambda=-\sum_{\{i,j\}\in\Lambda_b}\sum_{\mu=\uparrow,\downarrow}
\left(t_{i,j}c_{i,\mu}^\dagger c_{j,\mu}+t_{i,j}^\ast c_{j,\mu}^\dagger c_{i,\mu}
\right)+V(\{n_{i,\mu}\})+\sum_{i\in\Lambda_s}{\bf B}_i\cdot{\bf S}_i,
\label{hamelectron}
\end{equation}
where $c_{i,\mu}^\dagger, c_{i,\mu}$ are, respectively, 
the electron creation and annihilation operators with the $z$ component 
of the spin $\mu=\uparrow,\downarrow$, $n_{i,\mu}=c_{i,\mu}^\dagger c_{i,\mu}$ 
is the corresponding number operator, and ${\bf S}_i=(S_i^x,S_i^y,S_i^z)$ 
are the spin operator given by 
$S_i^a=\sum_{\mu,\nu=\uparrow,\downarrow}c_{i,\mu}^\dagger \sigma_{\mu,\nu}^a
c_{i,\nu}$ with the Pauli spin matrix $(\sigma_{\mu,\nu}^a)$ for $a=x,y,z$; 
$t_{i,j}\in{\bf C}$ are the hopping amplitude, $V(\{n_{i,\mu}\})$ is a real 
function of the number operators, and ${\bf B}_i=(B_i^x,B_i^y,B_i^z)\in{\bf R}^3$ 
are local magnetic fields. We assume that the interaction $V(\{n_{i,\mu}\})$ is  
of finite range in the sense of the graph theoretic distance.   

Clearly the Hamiltonian $H_\Lambda$ of (\ref{hamelectron}) commutes with the total number 
operator ${\cal N}_\Lambda=\sum_{i\in\Lambda_s}\sum_{\mu=\uparrow,\downarrow}n_{i,\mu}$ 
for a finite volume $|\Lambda_s|<\infty$. 
We denote by $H_{\Lambda,N}$ the restriction of $H_\Lambda$ onto 
the eigenspace of ${\cal N}_\Lambda$ with the eigenvalue $N$. 
Let $P_{0,N}$ be the projection onto the sector of the ground state 
of $H_{\Lambda,N}$, and we denote the ground-state expectation by 
\begin{equation}
\left\langle\cdots\right\rangle_{0,N}=\frac{1}{q_N}
{\rm Tr}\ (\cdots)P_{0,N},  
\end{equation}
where $q_N$ is the degeneracy of the ground state. 
Assume the existence of a uniform gap above the ground state. Then we have 
\begin{equation}
\left|\left\langle c_{m,\mu}^\dagger c_{n,\mu}\right\rangle_{0,N}\right|\le
{\rm Const.}\exp\left[-\gamma\{{\rm dist}(m,n)\}^{1-D/2}
\right] 
\end{equation}
and 
\begin{equation}
\left|\left\langle c_{m,\uparrow}^\dagger c_{m,\downarrow}^\dagger 
c_{n,\uparrow}c_{n,\downarrow}\right\rangle_{0,N}\right|
\le{\rm Const.}\exp\left[-2\gamma\{{\rm dist}(m,n)\}^{1-D/2}\right] 
\end{equation}
with some constant $\gamma$ in the thermodynamic limit $|\Lambda_s|\rightarrow\infty$.
If the local magnetic field has the form ${\bf B}_i=(0,0,B_i)$, then 
we further have 
\begin{equation}
\left|\left\langle S_m^+ S_n^-\right\rangle_{0,N}\right|\le
{\rm Const.}\exp\left[-\gamma'\{{\rm dist}(m,n)\}^{1-D/2}\right] 
\end{equation} 
with some constant $\gamma'$. 
We remark that, in the latter situation, we can also restrict 
the Hamiltonian $H_{\Lambda,N}$ and the expectation 
to the subspace with a fixed total magnetization.

\Section{Proof of the main theorems}
\label{proof}

We will give a proof only for Theorem~\ref{maintheorem} 
because the rest of the bounds can be proved in the same way. 
The application to other systems is also straightforward. 

Before proceeding to the proof, let us sketch the idea of the 
proof and the key tools. 
In the previous work \cite{KT1}, the global quantum mechanical U(1) symmetry was 
used for estimating the correlation functions of the Hubbard model at 
finite temperatures. Roughly speaking, the strength 
of the long-range ordering can be measured 
by twisting the U(1) phase locally in the pure 
imaginary direction. The difference between zero temperature and 
finite temperatures is in their density matrices. 
Since the projection $P_{0,M}$ onto the sector of the ground state 
can be written in the contour integral of the resolvent,  
we apply the method of \cite{KT1} to the resolvent $(z-H_\Lambda)^{-1}$ 
instead of the Boltzmann weight for finite temperatures. 
It is well known that, for a single-particle resolvent with a classically forbidden 
energy, the Combes-Thomas method \cite{CT} 
yields the WKB-type tunnelling estimate, i.e., the exponentially decaying bound. 
In the method, twisting the quantum mechanical phase locally 
in the pure imaginary direction plays an important role, too, 
and so the Combes-Thomas method is essentially equivalent to the McBryan-Spencer method.  
In the present paper, we use the improved version \cite{BCH} of the Combes-Thomas 
method. In order to extend the Combes-Thomas method to many-body systems in 
statistical mechanics, 
we further employ the technique \cite{HvL,KomaTasaki} which was 
developed for many-body systems to treat quantities of order of the volume.  

In order to estimate the transverse spin-spin correlation, 
$\left\langle S_m^+S_n^-\right\rangle_{0,M}$, we introduce ``gauge transformation", 
\begin{equation} 
G(\alpha):=\prod_{i\in\Lambda_s}\exp[\alpha\theta_i S_i^z],
\label{defGalpha}
\end{equation}
where $\alpha$ is a real parameter to be determined.  
For the choice of the function $\theta_i$, we follow Picco's idea \cite{Picco}, 
but we modify it in order to obtain 
a better decay bound for the correlation. 
Let $\kappa$ be a positive parameter satisfying 
\begin{equation}
1-\frac{D}{2}<\kappa<\frac{3}{2}-\frac{D}{2},
\label{kappa}
\end{equation}
and write $R={\rm dist}(n,m)$.
We choose the real function $\theta_i$ on the lattice $\Lambda_s$ as  
\begin{equation}
\theta_\ell=R^{1-D/2}\times\cases{R^{-\kappa}-1, & for $\ell=m$;\cr
       [{\rm dist}(\ell,m)/R]^\kappa-1, & for $1\le{\rm dist}(\ell,m)\le R$;\cr 
                    0, & for ${\rm dist}(\ell,m)>R$.}
\label{deftheta}
\end{equation}
 
Note that 
\begin{equation}
G(\alpha)^{-1}S_i^\pm G(\alpha)=e^{\pm\alpha\theta_i}S_i^\pm 
\quad\mbox{for }\ i\in\Lambda_s.
\label{transSpm}
\end{equation}
Using the relation (\ref{transSpm}), we have 
\begin{eqnarray}
{\rm Tr}\ S_m^+S_n^-P_{0,M}&=&{\rm Tr}\ G(\alpha)^{-1}S_m^+G(\alpha)G(\alpha)^{-1}
S_n^+G(\alpha)G(\alpha)^{-1}P_{0,M}G(\alpha)\ret
&=&e^{\alpha(\theta_m-\theta_n)}\ {\rm Tr}\ S_m^+S_n^-P_{0,M}(\alpha),
\label{corrGtrans}
\end{eqnarray}
where we have written as 
\begin{equation}
P_{0,M}(\alpha):=G(\alpha)^{-1}P_{0,M}G(\alpha).
\label{defPalpha}
\end{equation}
This operator has the property, 
\begin{eqnarray}
P_{0,M}(\alpha)^2&=&G(\alpha)^{-1}P_{0,M}G(\alpha)G(\alpha)^{-1}P_{0,M}G(\alpha)\ret
&=&G(\alpha)^{-1}P_{0,M}^2G(\alpha)\ret
&=&P_{0,M}(\alpha). 
\end{eqnarray}
Using this, we have 
\begin{eqnarray} 
\left|{\rm Tr}\ S_m^+S_n^-P_{0,M}(\alpha)\right|
&=&\left|{\rm Tr}\ S_m^+S_n^-P_{0,M}(\alpha)P_{0,M}(\alpha)\right|\ret
&\le&\sqrt{{\rm Tr}\ P_{0,M}(\alpha)^\ast S_n^+S_m^-S_m^+S_n^-P_{0,M}(\alpha)
\cdot{\rm Tr}\ P_{0,M}(\alpha)^\ast P_{0,M}(\alpha)}\ret
&\le&\left\Vert S_m^+S_n^-\right\Vert{\rm Tr}\ P_{0,M}(\alpha)^\ast 
P_{0,M}(\alpha).
\label{CorrTRbound} 
\end{eqnarray}
{From} the definitions, (\ref{defGalpha}) and (\ref{defPalpha}), one has 
\begin{eqnarray}
{\rm Tr}\ P_{0,M}(\alpha)^\ast P_{0,M}(\alpha)
&=&{\rm Tr}\ G(\alpha)P_{0,M}G(\alpha)^{-1}G(\alpha)^{-1}P_{0,M}G(\alpha)\ret
&=&{\rm Tr}\ G(\alpha)^{-2}P_{0,M}G(\alpha)^2P_{0,M}\ret
&\le&q_M\left\Vert P_{0,M}(2\alpha)\right\Vert. 
\end{eqnarray}
Combining this, (\ref{deftheta}), (\ref{corrGtrans}) 
and (\ref{CorrTRbound}), we have 
\begin{equation}
\left|\left\langle S_m^+S_n^-\right\rangle_{0,M}\right|
\le\left\Vert S_m^+S_n^-\right\Vert
\left\Vert P_{0,M}(2\alpha)\right\Vert
\exp\left[-\alpha\left(1-R^{-\kappa}\right)
\{{\rm dist}(m,n)\}^{1-D/2}\right].
\label{corrbound0}
\end{equation}
Therefore it is sufficient to evaluate the norm 
$\left\Vert P_{0,M}(2\alpha)\right\Vert$ with 
a suitable choice of the parameter $\alpha$. 

Let us introduce the contour integral representation of 
the projection $P_{0,M}$ as 
\begin{equation}
P_{0,M}=\frac{1}{2\pi i}\int_\Gamma\frac{dz}{z-H_\Lambda}P_M,
\label{P0Mintegral}
\end{equation}
where the closed path $\Gamma$ is taken to encircle all of the eigenvalues in 
the sector of the corresponding ground state, 
and $P_M$ is the projection onto the eigenspace ${\cal H}_{\Lambda,M}$ 
of $S_\Lambda^z$ with the eigenvalue $M$.
{From} the definition (\ref{defPalpha}), one has 
\begin{eqnarray}
P_{0,M}(2\alpha)&=&G(2\alpha)^{-1}\frac{1}{2\pi i}
\int_\Gamma\frac{dz}{z-H_\Lambda}P_MG(2\alpha)\ret
&=&G(2\alpha)^{-1}\frac{1}{2\pi i}
\int_\Gamma\frac{dz}{z-H_\Lambda}G(2\alpha)P_M\ret
&=&\frac{1}{2\pi i}
\int_\Gamma\frac{dz}{z-H_\Lambda'}P_M
\label{P0M2alphaintegral}
\end{eqnarray}
with the non-hermitian matrix $H_\Lambda'=G(2\alpha)^{-1}H_\Lambda G(2\alpha)$. 
In order to evaluate the resolvent $(z-H_\Lambda')^{-1}$ 
in the right-hand side, we begin with 
getting the explicit form of the transformed Hamiltonian $H_\Lambda'$.  
Note that the Hamiltonian $H_\Lambda^{\rm XY}$ of (\ref{hamXY}) is written as 
\begin{equation}
H_\Lambda^{\rm XY}=\sum_{\{i,j\}\in\Lambda_b}J_{i,j}^{\rm XY}
(S_i^+S_j^-+S_i^-S_j^+).
\end{equation} 
Using the definition (\ref{deftheta}) of $\{\theta_i\}$ and 
the relations (\ref{transSpm}), one has 
\begin{eqnarray}
G(2\alpha)^{-1}H_\Lambda^{\rm XY}G(2\alpha)&=&
\sum_{\{i,j\}\in\Lambda_b}J_{i,j}^{\rm XY} 
\left[e^{2\alpha(\theta_i-\theta_j)}S_i^+S_j^-+e^{-2\alpha(\theta_i-\theta_j)}
S_i^-S_j^+\right]\ret
&=&H_\Lambda^{\rm XY}+K_\Lambda+iL_\Lambda
\end{eqnarray}
with the two hermitian matrices, 
\begin{equation}
K_\Lambda=\sum_{\{i,j\}\in A_{1,R}(m)}J_{i,j}^{\rm XY}
\left\{\cosh[2\alpha(\theta_i-\theta_j)]-1\right\}
\left(S_i^+S_j^-+S_i^-S_j^+\right)
\label{defK}
\end{equation}
and
\begin{equation}
L_\Lambda=-i\sum_{\{i,j\}\in A_{1,R}(m)}J_{i,j}^{\rm XY}
\sinh[2\alpha(\theta_i-\theta_j)]
\left(S_i^+S_j^--S_i^-S_j^+\right),
\end{equation}
where the set $A_{1,R}(m)$ of the bonds is given by 
\begin{equation}
A_{1,R}(m)=\left\{\{i,j\}\in\Lambda_b|1\le{\rm dist}(i,m)\le R-1, 
{\rm dist}(j,m)\ge{\rm dist}(i,m)\right\}. 
\end{equation}
Thus we have 
\begin{equation}
H_\Lambda'=H_\Lambda+K_\Lambda+iL_\Lambda.
\end{equation}

The norms of the operators, $K_\Lambda$ and $L_\Lambda$, can be estimated 
as follows: 

\begin{lemma}
\label{lemma:Kbound}
The norm of the operator $K_\Lambda$ satisfies 
\begin{equation}
\Vert K_\Lambda\Vert\le J_{\rm max}^{\rm XY}
\left\Vert\left(S_i^+S_j^-+S_i^-S_j^+\right)\right\Vert
C_0^2(\cosh2\alpha-1)\frac{2\kappa+D-1}{2\kappa+D-2}.
\label{Kbound}
\end{equation}
\end{lemma}

\noindent
{\bf Remark:} The bound implies that one can make 
the contribution from $K_\Lambda$ small 
in the resolvent $(z-H_\Lambda')^{-1}$ by choosing a small $\alpha$.

\begin{proof}{Proof}
{From} (\ref{defK}), one has 
\begin{equation}
\Vert K_\Lambda\Vert\le J_{\rm max}^{\rm XY}
\left\Vert\left(S_i^+S_j^-+S_i^-S_j^+\right)\right\Vert
\times\sum_{\{i,j\}\in A_{1,R}(m)}\{\cosh[2\alpha(\theta_i-\theta_j)]-1\}.
\label{Kcoshbound} 
\end{equation}
The sum in the right-hand side is rewritten as
\begin{equation}
\sum_{\{i,j\}\in A_{1,R}(m)}\{\cosh[2\alpha(\theta_i-\theta_j)]-1\}
=\sum_{r=1}^{R-1}\sum_{i:{\rm dist}(i,m)=r}
\mathop{\sum_{j:\{i,j\}\in\Lambda_b}}_{{\rm dist}(j,m)=r+1}
\{\cosh[2\alpha(\theta_i-\theta_j)]-1\}.
\label{coshbound} 
\end{equation} 
{From} the definitions (\ref{kappa}) and (\ref{deftheta}), one has  
\begin{equation}
|\theta_i-\theta_j|\le\kappa R^{-\kappa+1-D/2}r^{\kappa-1}\le 1
\end{equation}
for $i,j$ satisfying 
${\rm dist}(i,m)=r$ and ${\rm dist}(j,m)=r+1$ for $r=1,2,\ldots,R-1$. 
{From} the second inequality, 
\begin{equation}
\frac{\cosh[2\alpha(\theta_i-\theta_j)]-1}{4\alpha^2(\theta_i-\theta_j)^2}
\le\frac{\cosh2\alpha-1}{4\alpha^2}.
\end{equation}
Combining these inequalities, one obtains 
\begin{equation}
\cosh[2\alpha(\theta_i-\theta_j)]-1\le(\cosh2\alpha-1)(\theta_i-\theta_j)^2
\le(\cosh2\alpha-1)R^{-2\kappa+2-D}r^{2\kappa-2}.
\end{equation}
Substituting this into the right-hand side of (\ref{coshbound}), we have
\begin{eqnarray} 
& &\sum_{\{i,j\}\in A_{1,R}(m)}\{\cosh[2\alpha(\theta_i-\theta_j)]-1\}\ret
&\le&(\cosh2\alpha-1)\sum_{r=1}^{R-1}\sum_{i:{\rm dist}(i,m)=r}
\mathop{\sum_{j:\{i,j\}\in\Lambda_b}}_{{\rm dist}(j,m)=r+1}
R^{-2\kappa+2-D}r^{2\kappa-2}\ret
&\le&C_0(\cosh2\alpha-1)\sum_{r=1}^{R-1}\sum_{i:{\rm dist}(i,m)=r}
R^{-2\kappa+2-D}r^{2\kappa-2}\ret
&\le&C_0^2(\cosh2\alpha-1)\sum_{r=1}^{R-1}r^{D-1}R^{-2\kappa+2-D}r^{2\kappa-2}\ret
&\le&C_0^2(\cosh2\alpha-1)\frac{2\kappa+D-1}{2\kappa+D-2},
\end{eqnarray}
where we have used the assumption (\ref{dimensionD}) on the (fractal) dimension $D$ 
and the definition (\ref{kappa}) of the parameter $\kappa$. 
Substituting this into the right-hand side of (\ref{Kcoshbound}) gives 
the desired bound (\ref{Kbound}). 
\end{proof}

In a similar way, we can obtain the following bound for $\Vert L_\Lambda\Vert$:

\begin{lemma}
\label{lemma:Lnorm}
The norm of the operator $L_\Lambda$ satisfies 
\begin{equation}
\Vert L_\Lambda\Vert\le J_{\rm max}^{\rm XY}
\left\Vert\left(S_i^+S_j^--S_i^-S_j^+\right)\right\Vert
C_0^2|\sinh2\alpha|\left(1+\frac{R^{D/2}}{\kappa+D-1}\right).
\label{Lbound}
\end{equation}
\end{lemma}

\begin{proof}{Proof}
In the same way as in the proof of Lemma~\ref{lemma:Kbound}, one has 
\begin{equation}
\Vert L_\Lambda\Vert\le J_{\rm max}^{\rm XY}
\left\Vert\left(S_i^+S_j^--S_i^-S_j^+\right)\right\Vert
\times\sum_{r=1}^{R-1}\sum_{i:{\rm dist}(i,m)=r}
\mathop{\sum_{j:\{i,j\}\in\Lambda_b}}_{{\rm dist}(j,m)=r+1}
|\sinh[2\alpha(\theta_i-\theta_j)]|.
\label{Ksinhbound}
\end{equation}
For $\theta_i,\theta_j$ in the sum, the following bound is valid: 
\begin{equation}
\frac{|\sinh[2\alpha(\theta_i-\theta_j)]|}{|2\alpha(\theta_i-\theta_j)|}
\le\frac{|\sinh2\alpha|}{|2\alpha|}.
\end{equation}
Therefore we have 
\begin{equation}
|\sinh[2\alpha(\theta_i-\theta_j)]|\le|\sinh 2\alpha||\theta_i-\theta_j|
\le|\sinh 2\alpha|R^{-\kappa+1-D/2}r^{\kappa-1}.
\end{equation}
Substituting this into the right-hand side of (\ref{Ksinhbound}), 
we can obtain the desired bound (\ref{Lbound}) in the same way as in the proof 
of Lemma~\ref{lemma:Kbound}.
\end{proof}

The upper bound (\ref{Lbound}) increases as the distance $R$ between the two spins 
increases. In fact, this bound is not sufficient for estimating 
the resolvent $(z-H_\Lambda')^{-1}$. We will further employ the technique 
developed in \cite{HvL,KomaTasaki}. For this purpose, we need 
the following estimate for the double commutator:  

\begin{lemma}
\label{lemma:doublecommuL}
The following bound is valid:
\begin{equation}
\left\Vert[L_\Lambda,[H_\Lambda,L_\Lambda]]\right\Vert\le
C_1(\sinh2\alpha)^2\frac{2\kappa+D-1}{2\kappa+D-2},
\label{doublecommuestimate} 
\end{equation}
where $C_1$ is a positive constant which is independent of 
the parameter $\alpha$.  
\end{lemma}

\begin{proof}{Proof}
Write the Hamiltonian $H_\Lambda$ in terms of the local Hamiltonian $h_{u,v}$ as 
\begin{equation}
H_\Lambda=\sum_{\{u,v\}\in\Lambda_b}h_{u,v}
\quad\mbox{with }\ h_{u,v}=J_{u,v}^{\rm XY}(S_u^+S_v^-+S_u^-S_v^+)
+J_{u,v}^{\rm Z}S_u^zS_v^z. 
\end{equation}
Note that 
\begin{eqnarray}
[H_\Lambda,L_\Lambda]&=&-i\sum_{\{u,v\}\in\Lambda_b}
\sum_{\{i,j\}\in A_{1,R}(m)}J_{i,j}^{\rm XY}\sinh[2\alpha(\theta_i-\theta_j)]
\left[h_{u,v},(S_i^+S_j^--S_i^-S_j^+)\right]\ret
&=&-i\sum_{\{i,j\}\in A_{1,R}(m)}J_{i,j}^{\rm XY}\sinh[2\alpha(\theta_i-\theta_j)]
\sum_{\{u,v\}\cap\{i,j\}\ne\emptyset}\left[h_{u,v},(S_i^+S_j^--S_i^-S_j^+)\right]\ret
&=&-i\sum_{\{i,j\}\in A_{1,R}(m)}J_{i,j}^{\rm XY}\sinh[2\alpha(\theta_i-\theta_j)]
\sum_{t\in\Lambda_s:{\rm dist}(t,\{i,j\})=0,1}M_{i,j;t},
\end{eqnarray}
where $M_{i,j;t}$ is a matrix with the support $\{i,j,t\}\subset\Lambda_s$. 
Using this, the double commutator is written as 
\begin{eqnarray}
\left[L_\Lambda,[H_\Lambda,L_\Lambda]\right]&=&
-\sum_{\{k,\ell\}\in A_{1,R}(m)}\sum_{\{i,j\}\in A_{1,R}(m)}
J_{k,\ell}^{\rm XY}J_{i,j}^{\rm XY}\sinh[2\alpha(\theta_k-\theta_\ell)]
\sinh[2\alpha(\theta_i-\theta_j)]\ret
&\times&\sum_{t\in\Lambda_s:{\rm dist}(t,\{i,j\})=0,1}
\left[(S_k^+S_\ell^--S_k^-S_\ell^+),M_{i,j;t}\right].
\end{eqnarray}
In the same way as in the proof of Lemma~\ref{lemma:Kbound}, we have 
\begin{eqnarray}
\left\Vert\left[L_\Lambda,[H_\Lambda,L_\Lambda]\right]\right\Vert&\le&
\left(J_{\rm max}^{\rm XY}\right)^2(\sinh2\alpha)^2R^{-2\kappa+2-D}\ret
&\times&\sum_{r'=1}^{R-1}
\sum_{k:{\rm dist}(k,m)=r'}\mathop{\sum_{\{k,\ell\}\in\Lambda_b}}_{\ell:
{\rm dist}(\ell,m)=r'+1}
\sum_{r=1}^{R-1}
\sum_{i:{\rm dist}(i,m)=r}\mathop{\sum_{\{i,j\}\in\Lambda_b}}_{j:
{\rm dist}(j,m)=r+1}(r')^{\kappa-1}r^{\kappa-1}\ret
&\times&\sum_{t\in\Lambda_s:{\rm dist}(t,\{i,j\})=0,1}
\left\Vert\left[(S_k^+S_\ell^--S_k^-S_\ell^+),M_{i,j;t}\right]\right\Vert.
\label{doublecommubound}
\end{eqnarray}
We decompose the sum in the right-hand side 
into two parts, $I_1$ with $r>r'$ and $I_2$ with $r'\ge r$, as 
\begin{eqnarray}
I_1&=&\sum_{r'=1}^{R-1}
\sum_{k:{\rm dist}(k,m)=r'}\mathop{\sum_{\{k,\ell\}\in\Lambda_b}}_{\ell:
{\rm dist}(\ell,m)=r'+1}
\sum_{r>r'}^{R-1}
\sum_{i:{\rm dist}(i,m)=r}\mathop{\sum_{\{i,j\}\in\Lambda_b}}_{j:
{\rm dist}(j,m)=r+1}(r')^{\kappa-1}r^{\kappa-1}\ret
&\times&\sum_{t\in\Lambda_s:{\rm dist}(t,\{i,j\})=0,1}
\left\Vert\left[(S_k^+S_\ell^--S_k^-S_\ell^+),M_{i,j;t}\right]\right\Vert
\label{defI1}
\end{eqnarray}
and 
\begin{eqnarray}
I_2&=&
\sum_{r=1}^{R-1}
\sum_{i:{\rm dist}(i,m)=r}\mathop{\sum_{\{i,j\}\in\Lambda_b}}_{j:
{\rm dist}(j,m)=r+1}
\sum_{r'\ge r}^{R-1}
\sum_{k:{\rm dist}(k,m)=r'}\mathop{\sum_{\{k,\ell\}\in\Lambda_b}}_{\ell:
{\rm dist}(\ell,m)=r'+1}
(r')^{\kappa-1}r^{\kappa-1}\ret
&\times&\sum_{t\in\Lambda_s:{\rm dist}(t,\{i,j\})=0,1}
\left\Vert\left[(S_k^+S_\ell^--S_k^-S_\ell^+),M_{i,j;t}\right]\right\Vert.
\label{defI2}
\end{eqnarray}

First let us estimate $I_2$. Since $r'\ge r$ in the sum, 
one has $(r')^{\kappa-1}\le r^{\kappa-1}$. Using this inequality, $I_2$ 
is evaluated as  
\begin{eqnarray}
I_2&\le&
\sum_{r=1}^{R-1}
\sum_{i:{\rm dist}(i,m)=r}\mathop{\sum_{\{i,j\}\in\Lambda_b}}_{j:
{\rm dist}(j,m)=r+1}r^{2\kappa-2}\ret
&\times&\sum_{r'\ge r}^{R-1}
\sum_{k:{\rm dist}(k,m)=r'}\mathop{\sum_{\{k,\ell\}\in\Lambda_b}}_{\ell:
{\rm dist}(\ell,m)=r'+1}
\sum_{t\in\Lambda_s:{\rm dist}(t,\{i,j\})=0,1}
\left\Vert\left[(S_k^+S_\ell^--S_k^-S_\ell^+),M_{i,j;t}\right]\right\Vert\ret
&\le&\sum_{r=1}^{R-1}
\sum_{i:{\rm dist}(i,m)=r}\mathop{\sum_{\{i,j\}\in\Lambda_b}}_{j:
{\rm dist}(j,m)=r+1}r^{2\kappa-2}\ret
&\times&\sum_{t\in\Lambda_s:{\rm dist}(t,\{i,j\})=0,1}
\sum_{\{k,\ell\}\cap\{i,j,t\}\ne\emptyset}
\left\Vert\left[(S_k^+S_\ell^--S_k^-S_\ell^+),M_{i,j;t}\right]\right\Vert\ret
&\le&12C_0^4
\max\left\{\left\Vert\left[(S_k^+S_\ell^--S_k^-S_\ell^+),M_{i,j;t}\right]\right\Vert\right\}
\sum_{r=1}^{R-1}r^{D-1}r^{2\kappa-2}\ret
&\le&12C_0^4
\max\left\{\left\Vert\left[(S_k^+S_\ell^--S_k^-S_\ell^+),M_{i,j;t}\right]\right\Vert\right\}
\left[1+\frac{R^{2\kappa+D-2}}{2\kappa+D-2}\right].
\end{eqnarray}
Similarly we have the same upper bound for $I_1$. Combining these bounds, 
(\ref{doublecommubound}), (\ref{defI1}) and (\ref{defI2}), the desired result 
(\ref{doublecommuestimate}) is obtained. 
\end{proof}

Now let us estimate $P_{0,M}(2\alpha)$ of (\ref{P0M2alphaintegral}). 
The contour integral in the right-hand side is written 
\begin{eqnarray}
\int_\Gamma\frac{dz}{z-H_\Lambda'}P_M&=&
\int_{-y_0}^{y_0}\frac{idy}{E_++iy-H_\Lambda'}P_M
+\int_{E_+}^{E_-}\frac{dx}{x+iy_0-H_\Lambda'}P_M\ret 
&+&\int_{y_0}^{-y_0}\frac{idy}{E_-+iy-H_\Lambda'}P_M
+\int_{E_-}^{E_+}\frac{dx}{x-iy_0-H_\Lambda'}P_M.
\label{P0M2alphaintegral2}
\end{eqnarray}
Here we choose the three real numbers, $y_0, E_+$ and $E_-$, as follows: 
Relying on Lemma~\ref{lemma:Lnorm}, we take 
\begin{equation}
y_0=C_2R^{D/2}
\label{defy0} 
\end{equation}
satisfying   
\begin{equation}
C_2R^{D/2}-\Vert L_\Lambda\Vert\ge C_3>0
\label{y0condition}
\end{equation}
with some positive constants $C_2$ and $C_3$. 
{From} the assumption on the spectrum of the Hamiltonian $H_\Lambda$, 
we can take $E_+,E_-$ so that the distance between the spectrum and 
$\{E_+,E_-\}$ is greater than or equal to $\Delta E/2$. 
{From} (\ref{P0M2alphaintegral2}) and (\ref{defy0}), we have 
\begin{eqnarray}
\Vert P_{0,M}(2\alpha)\Vert&\le& 
\frac{C_2}{\pi}R^{D/2}
\left\{\sup_{y\in[-y_0,y_0]}\left\Vert R'(E_++iy)P_M\right\Vert
+\sup_{y\in[-y_0,y_0]}\left\Vert R'(E_-+iy)P_M\right\Vert\right\}\ret
&+&\frac{E_+-E_-}{2\pi}\left\{\sup_{x\in[E_-,E_+]}\left\Vert R'(x+iy_0)
\right\Vert+\sup_{x\in[E_-,E_+]}\left\Vert R'(x-iy_0)
\right\Vert\right\},
\label{P0M2alphanormbound}
\end{eqnarray}
where we have written $R'(z)=(H_\Lambda'-z)^{-1}$. 

In order to estimate the norm of the resolvent $R'(z)$ with $z=E_++iy$, 
we employ the technique developed in \cite{HvL,KomaTasaki} for 
the following matrix element $\left\langle\Phi_+,L_\Lambda\Phi_-\right\rangle$: 

\begin{lemma}
\label{lemma:matrixelementLPhipm}
Let $\Phi_-=P_{0,M}\Phi$ and 
$\Phi_+=(1-P_{0,M})\Phi$ for a vector $\Phi\in{\cal H}_{\Lambda,M}$. 
Then we have 
\begin{equation}
\left|\left\langle\Phi_+,L_\Lambda\Phi_-\right\rangle\right|
\le f(\alpha,\Lambda)\Vert\Phi_+\Vert\Vert\Phi_-\Vert,
\label{LPhipmbound}
\end{equation}
where 
\begin{equation}
f(\alpha,\Lambda)=\sqrt{\frac{1}{2\Delta E}\Vert[L_\Lambda,[H_\Lambda,L_\Lambda]]\Vert
+2\frac{\Delta{\cal E}}{\Delta E}\Vert L_\Lambda\Vert^2}
\label{falphaLambda}
\end{equation}
with $\Delta {\cal E}=\max_{\mu,\mu'}\{|E_{0,\mu}-E_{0,\mu'}|\}$.
\end{lemma}

\begin{proof}{Proof}
Note that 
\begin{eqnarray}
\left|\left\langle\Phi_+,L_\Lambda\Phi_-\right\rangle\right|^2&=&
\left\langle\Phi_-,L_\Lambda\Phi_+\right\rangle
\left\langle\Phi_+,L_\Lambda\Phi_-\right\rangle\ret
&\le&\Vert\Phi_+\Vert^2\left\langle\Phi_-,L_\Lambda(1-P_{0,M})
L_\Lambda\Phi_-\right\rangle.
\label{LPhipm2bound}
\end{eqnarray}
The matrix element in the right-hand side is evaluated as   
\begin{eqnarray}
\left\langle\Phi_-,L_\Lambda(1-P_{0,M})
L_\Lambda\Phi_-\right\rangle&\le&\frac{1}{\Delta E}
\left\langle\Phi_-,L_\Lambda(H_\Lambda-{\overline E}_0)(1-P_{0,M})
L_\Lambda\Phi_-\right\rangle\ret
&=&\frac{1}{\Delta E}
\left\langle\Phi_-,L_\Lambda(H_\Lambda-{\overline E}_0)
L_\Lambda\Phi_-\right\rangle\ret
&-&\frac{1}{\Delta E}
\left\langle\Phi_-,L_\Lambda(H_\Lambda-{\overline E}_0)P_{0,M}
L_\Lambda\Phi_-\right\rangle\ret
&\le&\frac{1}{\Delta E}
\left\langle\Phi_-,L_\Lambda(H_\Lambda-{\overline E}_0)
L_\Lambda\Phi_-\right\rangle+\frac{\Delta{\cal E}}{\Delta E}\Vert L_\Lambda\Vert^2
\Vert\Phi_-\Vert^2,\ret
\label{PhiL(1-P)LPhi}
\end{eqnarray}
where we have written ${\overline E}_0=\sum_{\mu=1}^{q_M}E_{0,\mu}/q_M$. 
Further the first term in the right-hand side in the last line 
can be written  
\begin{eqnarray}
\left\langle\Phi_-,L_\Lambda(H_\Lambda-{\overline E}_0)
L_\Lambda\Phi_-\right\rangle&=&
\left\langle\Phi_-,L_\Lambda[H_\Lambda,L_\Lambda]\Phi_-\right\rangle
+\left\langle\Phi_-,L_\Lambda^2(H_\Lambda-{\overline E}_0)\Phi_-\right\rangle.\ret
\end{eqnarray}
Therefore the matrix element can be evaluated as  
\begin{eqnarray}
& &\left|\left\langle\Phi_-,L_\Lambda(H_\Lambda-{\overline E}_0)
L_\Lambda\Phi_-\right\rangle\right|\ret
&\le&
\frac{1}{2}\left|\left\langle\Phi_-,L_\Lambda[H_\Lambda,L_\Lambda]\Phi_-\right\rangle
+\left\langle L_\Lambda[H_\Lambda,L_\Lambda]\Phi_-,\Phi_-\right\rangle\right|
+\Delta{\cal E}\Vert L_\Lambda\Vert^2\Vert\Phi_-\Vert^2\ret
&=&\frac{1}{2}\left|\left\langle\Phi_-,
[L_\Lambda[H_\Lambda,L_\Lambda]]\Phi_-\right\rangle\right|
+\Delta{\cal E}\Vert L_\Lambda\Vert^2\Vert\Phi_-\Vert^2\ret
&\le&\left(\frac{1}{2}\left\Vert[L_\Lambda[H_\Lambda,L_\Lambda]]\right\Vert
+\Delta{\cal E}\Vert L_\Lambda\Vert^2\right)\Vert\Phi_-\Vert^2. 
\end{eqnarray}
Combining this, (\ref{LPhipm2bound}) and (\ref{PhiL(1-P)LPhi}) gives 
the bound (\ref{LPhipmbound}) with (\ref{falphaLambda}).
\end{proof}

By using this lemma and the improved Combes-Thomas method \cite{BCH}, 
we obtain the following lemma:  

\begin{lemma}
\label{lemma:RE+}
Let $z=E_++iy$ with $y\in{\bf R}$. For a sufficiently large volume 
$|\Lambda_s|$, there exist positive constant, $\alpha_0$ and $C_4$, such that 
\begin{equation}
\left\Vert(H_\Lambda'-z)^{-1}P_M\right\Vert\le C_4
\quad\mbox{for any }\ \alpha\le\alpha_0.
\end{equation}
Both of the constants, $\alpha_0$ and $C_4$, are independent of 
the volume $|\Lambda_s|$ and of the distance $R$ between the two spin operators, 
$S_m^+,S_n^-$. 
\end{lemma}

\begin{proof}{Proof}
Using the Schwarz inequality, one has 
\begin{equation}
\Vert\Phi\Vert\left\Vert(H_\Lambda'-z)\Phi\right\Vert
\ge{\rm Re}\left\langle(\Phi_+-\Phi_-),(H_\Lambda'-z)(\Phi_++\Phi_-)\right\rangle
\label{H'-zbound}
\end{equation}
for any vector $\Phi\in{\cal H}_{\Lambda,M}$,
where $\Phi_+=(1-P_{0,M})\Phi$ and $\Phi_-=P_{0,M}\Phi$.  

We recall the expression, $H_\Lambda'=H_\Lambda+K_\Lambda+iL_\Lambda$. 
For the hermitian part of $H_\lambda'-z$, one has  
\begin{eqnarray}
& &{\rm Re}
\left\langle(\Phi_+-\Phi_-),(H_\Lambda+K_\Lambda-E_+)(\Phi_++\Phi_-)\right\rangle\ret
&\ge&(E_1-\Vert K_\Lambda\Vert-E_+)\Vert\Phi_+\Vert^2
+(E_+-\Vert K_\Lambda\Vert-\max_\mu\{E_{0,\mu}\})\Vert\Phi_-\Vert^2\ret
&+&{\rm Re}\left(\left\langle\Phi_+,K_\Lambda\Phi_-\right\rangle
-\left\langle\Phi_-,K_\Lambda\Phi_+\right\rangle\right),
\end{eqnarray}
where $E_1$ is the energy of the first excited state, and 
$E_{0,\mu}$ are the eigenvalues of the ground state.  
Since the last term in the right-hand side is equal to zero, one has 
\begin{equation}
{\rm Re}
\left\langle(\Phi_+-\Phi_-),(H_\Lambda+K_\Lambda-E_+)(\Phi_++\Phi_-)\right\rangle
\ge\left(\frac{1}{2}\Delta E-\Vert K_\Lambda\Vert\right)\Vert\Phi\Vert^2,
\label{RematrixHam}
\end{equation}
where we have used the fact that the distance between the spectrum of $H_{\Lambda,M}$ 
and the energy $E_+$ is greater than or equal to $\Delta E/2$. 

For the rest of $H_\Lambda'-z$, one has  
\begin{eqnarray}
{\rm Re}\left\langle(\Phi_+-\Phi_-),(iL_\Lambda-iy)(\Phi_++\Phi_-)\right\rangle
&=&-{\rm Im}\left\langle(\Phi_+-\Phi_-),(L_\Lambda-y)(\Phi_++\Phi_-)\right\rangle\ret
&=&-{\rm Im}\left(\left\langle\Phi_+,L_\Lambda\Phi_-\right\rangle
-\left\langle\Phi_-,L_\Lambda\Phi_+\right\rangle\right)\ret
&=&-2\ {\rm Im}\left\langle\Phi_+,L_\Lambda\Phi_-\right\rangle. 
\label{ImmatrixelementL}
\end{eqnarray}
{From} (\ref{LPhipmbound}), (\ref{H'-zbound}), (\ref{RematrixHam}) 
and (\ref{ImmatrixelementL}), one obtains
\begin{eqnarray}
\Vert\Phi\Vert\Vert(H_\Lambda'-z)\Phi\Vert&\ge&
\left(\frac{1}{2}\Delta E-\Vert K_\Lambda\Vert\right)\Vert\Phi\Vert^2
-2f(\alpha,\Lambda)\Vert\Phi_+\Vert\Vert\Phi_-\Vert\ret
&=&\left[\frac{1}{2}\Delta E-\Vert K_\Lambda\Vert-f(\alpha,\Lambda)\right]
\Vert\Phi\Vert^2+f(\alpha,\Lambda)\left(\Vert\Phi_+\Vert-\Vert\Phi_-\Vert\right)^2\ret
&\ge&\left[\frac{1}{2}\Delta E-\Vert K_\Lambda\Vert-f(\alpha,\Lambda)\right]
\Vert\Phi\Vert^2.
\label{H'-zlowerbound} 
\end{eqnarray}
{From} the assumption (\ref{defDeltacalE}) on the quasidegeneracy of the ground state, 
Lemma~\ref{lemma:doublecommuL} and the expression (\ref{falphaLambda}) 
of $f(\alpha,\Lambda)$, one can find that $f(\alpha,\Lambda)$ becomes small 
for a small parameter $\alpha$, and for a sufficiently large volume $|\Lambda_s|$ 
compared to the distance $R$ between the two spin operators. 
Combining this observation with Lemma~\ref{lemma:Kbound}, we have that 
there exist positive constants, $\alpha_0$ and ${\tilde C}_4$, such that 
\begin{equation}
\frac{1}{2}\Delta E-\Vert K_\Lambda\Vert-f(\alpha,\Lambda)\ge {\tilde C}_4
\end{equation}
for any $\alpha\le\alpha_0$, and 
for a sufficiently large volume $|\Lambda_s|$ compared to the distance $R$ 
between the two spin operators. Substituting this inequality into 
the right-hand side of the above bound (\ref{H'-zlowerbound}), we obtain 
\begin{equation}
\Vert(H_\Lambda'-z)\Phi\Vert\ge {\tilde C}_4\Vert\Phi\Vert.
\end{equation}
Choosing $\Phi=(H_\Lambda'-z)^{-1}P_M\Psi$ with a vector $\Psi$, we have 
\begin{equation}
\Vert P_M\Psi\Vert\ge{\tilde C}_4\left\Vert(H_\Lambda'-z)^{-1}P_M\Psi\right\Vert.
\end{equation}
\end{proof}

Similarly one can obtain the following two lemmas: 

\begin{lemma}
\label{lemma:RE-}
Let $z=E_-+iy$ with $y\in{\bf R}$. There exists a positive constant $C_5$ 
such that
\begin{equation}
\left\Vert(H_\Lambda'-z)^{-1}P_M\right\Vert\le C_5
\quad\mbox{for any }\ \alpha\le\alpha_0,
\end{equation}
where $\alpha_0$ is the same as in the preceding lemma. 
\end{lemma}

\begin{proof}{Proof}
Using the Schwarz inequality, one has 
\begin{eqnarray}
\Vert\Phi\Vert\left\Vert(H_\Lambda'-z)\Phi\right\Vert&\ge&
{\rm Re}\left\langle\Phi,(H_\Lambda'-z)\Phi\right\rangle\ret
&\ge&\left(\min_\mu\{E_{0,\mu}\}-\Vert K_\Lambda\Vert-E_-\right)\Vert\Phi\Vert^2\ret
&\ge&\left(\frac{1}{2}\Delta E-\Vert K_\Lambda\Vert\right)\Vert\Phi\Vert^2
\end{eqnarray}
for any vector $\Phi\in{\cal H}_{\Lambda,M}$. Here we have used 
$\min_\mu\{E_{0,\mu}\}-E_-\ge\Delta E/2$. Therefore, in the same way as 
in the proof of the preceding lemma, one can prove the statement of the lemma. 
\end{proof}

\begin{lemma} 
\label{lemma:Rxpmiy}
Let $z=x\pm iy_0$ with $x\in{\bf R}$. Then 
\begin{equation}
\left\Vert(H_\Lambda'-z)^{-1}\right\Vert\le C_3^{-1}.
\end{equation}
\end{lemma}
 
\begin{proof}{Proof}
Using the Schwarz inequality and the definition (\ref{defy0}) of $y_0$ 
with the condition (\ref{y0condition}), one has 
\begin{eqnarray}
\Vert\Phi\Vert\left\Vert(H_\Lambda'-z)\Phi\right\Vert&\ge&
\left|{\rm Im}\left\langle\Phi,(H_\Lambda'-z)\Phi\right\rangle\right|\ret
&=&\left|\left\langle\Phi,(L_\Lambda\mp y_0)\Phi\right\rangle\right|\ret
&\ge&\left|\left(y_0-\Vert L_\Lambda\Vert\right)\right|\Vert\Phi\Vert^2
\ge C_3\Vert\Phi\Vert^2
\end{eqnarray}
for any vector $\Phi$. Taking $\Phi=(H_\Lambda'-z)^{-1}\Psi$ with a vector $\Psi$, 
the desired bound can be obtained. 
\end{proof}

Combining the bound (\ref{P0M2alphanormbound}) 
with the three Lemmas~\ref{lemma:RE+}, \ref{lemma:RE-} and \ref{lemma:Rxpmiy}, we have 
\begin{equation}
\left\Vert P_{0,M}(2\alpha)\right\Vert\le CR^{D/2}+C'
\quad\mbox{for any }\ \alpha\le\alpha_0
\end{equation}
with the positive constants $C$ and $C'$. 
Substituting this into (\ref{corrbound0}) 
and choosing $\alpha=\alpha_0$, we obtain the bound (\ref{SScorrbound}) 
for the spin-spin correlation. 
\bigskip\bigskip

\noindent
{\bf Acknowledgements:} The author would like to thank Matthew B. Hastings, 
Bruno Nachtergaele and Hal Tasaki for useful conversations. 
\bigskip\bigskip


\end{document}